\documentclass[a4paper,fleqn,usenatbib]{mnras}

\usepackage{mathptmx}
\usepackage[T1]{fontenc}
\usepackage{ae,aecompl}

\usepackage{graphicx}	% Including figure files
\usepackage{amsmath}	% Advanced maths commands
\usepackage{amssymb}	% Extra maths symbols

\newcommand{\referee}{}	% per cm-squared

\title[Bi-modal Explosions of SNe Ia]{Type Ia Supernovae with Bi-Modal Explosions Are Common -- Possible Smoking Gun for Direct Collisions of White Dwarfs}

\author[Subo Dong et al.]{
Subo Dong,$^{1}$\thanks{E-mail: dongsubo@pku.edu.cn}
Boaz Katz,$^{2,3}$
Doron Kushnir$^{3}$
and Jose L. Prieto$^{4,5}$
\\
$^{1}$Kavli Institute for Astronomy and Astrophysics, Peking University, Yi He Yuan Road 5, Hai Dian District, Beijing 100871, China\\
$^{2}$Department of Particle Physics and Astrophysics, Weizmann Institute of Science, Rehovot 76100, Israel\\
$^{3}$Institute for Advanced Study, Einstein Drive, Princeton, NJ, 08540, USA\\
$^{4}$N\'ucleo de Astronom\'ia de la Facultad de Ingenier\'ia, Universidad Diego Portales, Av. Ej\'ercito 441, Santiago, Chile\\
$^{5}$Millennium Institute of Astrophysics, Santiago, Chile\\
}

\date{Accepted September 1, 2015. Received July 11, 2015; in original form July 11, 2015}

\pubyear{2015}

\begin{document}
\label{firstpage}
\pagerange{\pageref{firstpage}--\pageref{lastpage}}
\maketitle

\begin{abstract}
We discover clear doubly-peaked line profiles in 3 out of $\sim 20$ type Ia supernovae (SNe Ia) with high-quality nebular-phase spectra. The profiles are consistently present in three well-separated Co/Fe emission features. The two peaks are respectively blue-shifted and red-shifted relative to the host galaxies and are separated by $\sim 5000 {\rm km/s}$. The doubly-peaked profiles directly reflect a bi-modal velocity distribution of the radioactive $^{56}$Ni in the ejecta that powers the emission of these SNe.  Due to their random orientations, only a fraction of
SNe with intrinsically bi-modal velocity distributions will appear as doubly-peaked spectra. Therefore SNe with intrinsic bi-modality are likely common, especially among the SNe in the low-luminosity part on the Philips relation $(\Delta{m_{15}(B)}\gtrsim1.3;\,\sim 40\%$ of all SNe Ia). Such bi-modality is naturally expected from direct collisions of white dwarfs (WDs) due to the detonation of both WDs and is demonstrated in a 3D $0.64 M_{\odot}$-$0.64 M_{\odot}$ WD collision simulation. In the future, with a large sample of nebular spectra and a comprehensive set of numerical simulations, the collision model can be unambiguously tested as the primary channel for type Ia SNe, and the distribution of nebular line profiles will either be a smoking gun or rule it out.
\end{abstract}


%%%%%%%%%%%%%%%%%%%%%%%%%%%%%%%%%%%%%%%%%%%%%%%%%%

%%%%%%%%%%%%%%%%% BODY OF PAPER %%%%%%%%%%%%%%%%%%

\section{Introduction}

Type Ia supernovae (SNe Ia) are well-known cosmological ``standard candles'' thanks to a tight empirical correlation (the Phillips relation established by \citealt{phillips}) between intrinsic peak luminosities and post-peak brightness decline-rates ($\Delta{m_{15}(B)}$). SNe Ia are powered by the decay of $^{56}$Ni produced from the explosion of Carbon-Oxygen White Dwarfs (WDs), but the explosion mechanism is unknown. The two popular scenarios, single-degenerate (WD accretion exceeding the Chandrasekhar limit) and double-degenerate mergers (merger of two close WDs that spiral in due to gravitational radiation), have many theoretical and observational challenges \citep{hillebrandt2000typeIa,maoz13}. For both scenarios, a serious challenge is that a successful ignition of an explosive detonation has never been convincingly demonstrated.

Direct, head-on collisions of WDs would undoubtedly lead to successful explosions due to the strong shocks formed during the high-velocity impacts \citep{kushnir13,Rosswog2009cwd, Raskin2009oti,Raskin2010pdd,Hawley2012zip}, but they had long been thought  to only occur in dense stellar environments (globular clusters) and responsible for a negligible fraction of SNe Ia.
It was recently shown by \cite{katz2012rate} that the rate of direct collisions in common field triple systems may be as high as the SNe Ia rate. Previously, \cite{Thompson11} argued that the secular Lidov-Kozai mechanism \citep{lidov, kozai62} in triples might play an important role in WD-WD mergers via gravitational radiation to produce SNe Ia and speculated that collisions may sometimes occur. The high collision probability due to the non-secular corrections to the Lidov-Kozai mechanism obtained by \cite{katz2012rate} raised the possibility that the majority of SNe Ia result from collisions.
Supporting evidence was provided in \cite{kushnir13} in which high resolution numerical simulations of WD collisions reproduced several robust observational features of SNe Ia, especially establishing that the full range of $^{56}$Ni necessary for all SNe Ia across the Phillips relation, from the faint-end of SN 1991bg-like events to luminous 1991T-like events, can be produced by collisions of typical WDs.

Here we report that nebular spectra of some SNe Ia show double-peaked line profiles that suggest bimodal distributions of radioactive $^{56}$Ni in the ejecta. We find that 3 out of a sample of 20 SNe show this structure, implying a much larger fraction when viewing angle effects are taken into account. Using simulations we show that collision models of SNe Ia can produce structures similar to those observed due to the existence of two separate detonation seeds, providing new direct evidence for collisions as a significant SNe Ia channel. The bi-modal distributions can serve as a touchstone in testing SN Ia explosion scenarios in general.

\section{Discovery of Double-Peak Nebular Emission Features}

\subsection{The Sample}
Motivated by significantly non-spherical $^{56}$Ni distributions expected from WD collisions, we search for unusual nebular emission features in archival type Ia observations. In the nebular phase, the SN spectrum is emission-line dominated, and the line profiles directly probe the underlying velocity distribution of the emitting materials along the line of sight (LOS) due to the Doppler effect. By this time, $^{56}$Ni has completely decayed into $^{56}$Co and $^{56}$Fe, and the Co/Fe lines retain the velocity distribution of the original $^{56}$Ni. 

We systematically collect spectra of SNe Ia from the archival data. The main sources are the Berkeley Supernova Ia Program (BSNIP) {\citep{berkeley}},  the Center for Astrophysics Supernova Program {\citep{cfa},
Carnegie Supernova Project {\citep{csp} and the compilation from various sources by the Online Supernova Spectrum Database (SUSPECT). \footnote{SUSPECT: http://www.nhn.ou.edu/$\sim$suspect/; a public repository to download the spectra: WISeREP \citep{wiserep}.} We have also included the spectra of SN 2011fe taken by \cite{ben}. In the collection, there are 55 SNe with 155 spectra covering the wavelength range of interest (5000 - 7000 \AA\,in the rest frame) and with phase greater than 170 days, which we define in this work as nebular phase. We focus on $20$ SNe with the highest Signal-to-Noise-Ratio (SNR) nebular spectra.  Note 
that we exclude from the sample two SNe (SN 1986G and SN 2004bv) with a strong Na I D absorption feature at $\lambda\lambda 5890, 5896$\AA\,\, identified at earlier epochs since such a feature hinders clear identification of doubly-peak profile at $\sim 5900$\AA. It is also important to be cautious about spectra with over-subtraction H-$\alpha$ emission line at $6563$\AA\,\, from the host-galaxy, which may confuse the analysis of the feature at $\sim 6600$\AA.

\subsection{SN 2007on: a SNe Ia with Double-Peak Nebular Emission Features}

We identify three SNe with clear evidence of doubly-peaked velocity profiles. Figure 1 shows one example, SN 2007on, which exhibits clear doubly-peaked profiles for three well-spaced Fe and Co emission-line features.

\begin{figure}
\includegraphics[width=\columnwidth]{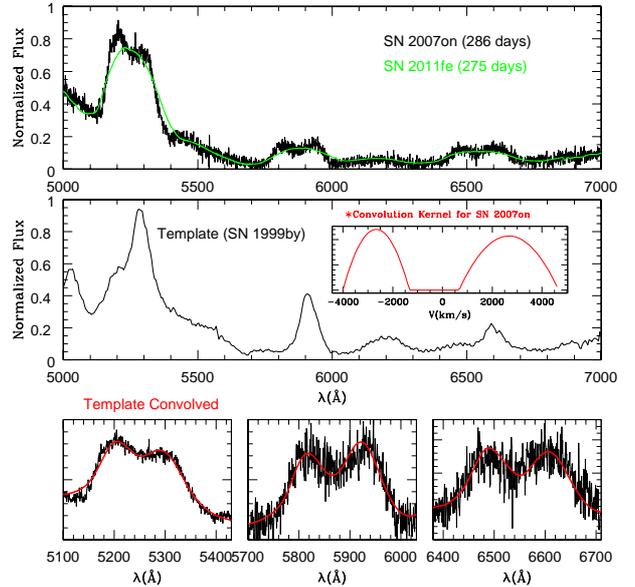}
\caption{The nebular spectrum of SN 2007on (top panel, black) has a clear doubly-peaked line profile appearing in three different Co/Fe emission features. The single-peak nebular spectrum of the ``normal'' SN 2011fe at a similar epoch is shown for comparison (green). The three double-peak profiles reflect the same bi-modal line-of-sight velocity distribution of the emitting Co/Fe materials. This is shown by fitting the spectrum with the convolution of a template spectrum using the same double-component velocity kernel (inset of middle panel, red). The template is chosen to be the narrow-width spectrum of SN 1999by (middle panel, black). The convolved spectrum (bottom panels, red) is compared with each of the three features of SN 2007on (bottom panels, black) by linear fitting that allows free normalizations and baseline flux shifts for each feature. The three features are due to blends of [FeIII] and [FeII] ($\sim5300${\AA}), [CoIII] ($\sim5900${\AA}), and blends of [CoIII] and [FeII] ($\sim6600${\AA}), respectively \citep{thesis,bower,turatto}. The $\sim 5900$ \AA\,\, [CoIII] feature is the most reliable among the three, and it is composed of two components ($5890$\AA\,\,and $5909$\AA) narrowly separated by $\sim 1000{\,\rm km/s}$, which is much smaller than the peak-to-peak separation of the double-peak profile $\sim 5000{\,\rm km/s}$.}
\end{figure}

\begin{figure}
\includegraphics[width=\columnwidth]{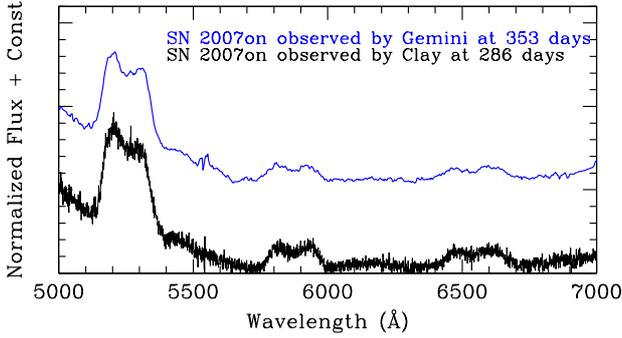}
\caption{The nebular spectra of SN 2007on observed by the Clay telescope at 286 days (black)
and the Gemini South telescope at 353 days (blue) show consistent doubly-peaked line profiles.}
\end{figure}

A general challenge in interpreting SNe line profiles is that many spectral features are due to blending from more than one line. A reasonable concern is that an observed doubly-peaked line profile could be due to two (or more) adjacent lines within an underlying single-peak velocity distribution. Several lines of evidence show that this is not the case for SN 2007on, so that the profile requires a bi-modal velocity distribution:

First, the doubly-peaked profiles occur for three widely-separated features emitted by two elements (Co and Fe) with the same peak-to-peak spacing $\Delta{\lambda}/\lambda \approx 0.02$. Furthermore, their shapes are all consistent with the same underlying velocity distribution. Figure 1 (bottom panel) shows that the spectra in three regions are well fitted by the convolution of a template spectrum with the same double-component velocity kernel (red line in the sub-panel of the middle panel). We choose the SN 1999by spectrum as the template (black line in the middle panel) due to its exceptionally narrow line width (similar to the famous SN 1991bg with SN 1999by having a higher SNR). A remarkable coincidence would be required in order that three additional lines conspire to produce such consistent profiles.

Next, other supernovae at a similar phase (thus a similar ratio between decaying Co and stable Fe) have similar line ratios among the three features but do not show the doubly-peaked profile. This is illustrated in the top panel of Figure 1 from comparison with SN 2011fe. While the line ratios of the two SNe are practically identical, the profiles of SN 2011fe are clearly single-peaked. The features from the same SN also show stable shapes in the nebular phase when multiple-epoch spectra have been taken.

Finally, the profile at $\sim 5800-6000$ \AA\,\,  is overwhelmingly dominated by one [CoIII] feature. %\footnote{The [CoIII] feature at $5900$ \AA\,\, is  composed of two narrowly separated components at $5890$ \AA\,\,  and $5909$ \AA\,\,, but their velocity separation $(\sim 1000{\,\rm km/s}$) is much smaller than the peak-to-peak separation of the double-line profile $(\sim 5000{\,\rm km/s})$}.
The line profile is thus ``clean'' and reliably reflects the underlying velocity distribution without the need of modelling. This is supported by nebular spectrum modeling \citep{thesis, bower} and re-affirmed by examining the spectra of SNe 1991bg and 1999by where the line widths are sufficiently narrow to allow for clear line identifications \citep{turatto}. The Co origin of the line has been firmly established for several SNe by examining the multiple-epoch spectra taken in the intervals 170-400 days, during which the Co decays significantly, and the lines maintain the same shape while the relative strengths with respect to Fe lines weaken according to the Co nuclear decay rate \citep{cobalt}.  The other two features at $\sim 5300$ \AA\,\, and $\sim 6600$ \AA\,\, are less reliable due to possible contributions from nearby lines. {\referee The frequently studied Fe feature at $\sim4700$  \AA\, shows significant blending from 
neighboring lines and is unsuitable (i.e., not clean enough) to be used for directly deriving the $^{56}$Ni velocity distribution from its profile. We examine
the choice of templates and spectral features in Appendix B}.
%{\referee As can be seen in Fig. B1, the nebular-phase spectra of SN 1991by and 
%SN 1991bg show good consistency in line shapes between 
% $\sim 5000$ and $7000$ \AA, supporting the use of spectral features in this 
%wavelength region to perform convolution. 
%In contrast, the spectra features in the range of $\sim4000 - 5000$ \AA\ 
%show significant difference in shape between SN 1999by and SN 1991bg, 
%in particular for the prominent Fe feature at $\sim4700$  \AA\, frequently 
%used to study SNe Ia nebular spectra.  The  $\sim4700$  \AA\ 
%feature has high ``shoulders'', indicating significant blending from 
%neighboring lines. Thus it is challenging to use the $\sim4700$ \AA\ 
%feature to directly derive the velocity distribution for a particular supernova.
%Nevertheless, in Appendix B., we examine the $\sim4700$ \AA\  features for the two 
%supernovae with bi-modal velocity distribution.}
%We find the description in {\cite{maeda}} a nebular-phase spectrum of
%SN 2007on taken at a different phase ($353 {\rm d}$) and telescope (Gemini) from that in our sample
%$286 {\rm d}$ with XXX). We retrieved archival late-time spectra of SN 2007on from the Gemini S%cience Archive (program GS-2008B-Q-8) and reduced it according to standard procedure. The r%educed spectrum is plotted in Figure 1. As can be seen, the double-peak features are clearly visible at this epoch, and they can be well described by the convolution of the double-peak
%profile used to fit the 286 d spectrum.

\citet{maeda} reported another nebular-phase spectrum of
SN 2007on taken at a different phase ($353 {\rm d}$) and telescope (Gemini South) from that in our sample
($286 {\rm d}$ at Clay).
We retrieved archival late-time spectra of SN 2007on from the Gemini Science Archive
(program GS-2008B-Q-8). The spectra were obtained with Gemini South + GMOS
on UT 2008-11-03 with the R150 grating and 1.0'' longslit. We reduced
the data using standard routines in the IRAF {\tt gemini.gmos} package,
including bias subtraction, flat-fielding, wavelength calibration using a CuAr lamp, 1D
spectral extraction, and flux calibration. We combined $4\times1200$~sec exposures
to obtain the final spectrum with wavelength coverage $4000-7500$~\AA\ and FWHM resolution of 22~\AA. The reduced spectrum is plotted in Figure 2. As can be seen, the double-peak features are clearly visible at this epoch, and they are consistent with those in the 286d spectrum.

\begin{figure}
\includegraphics[width=\columnwidth]{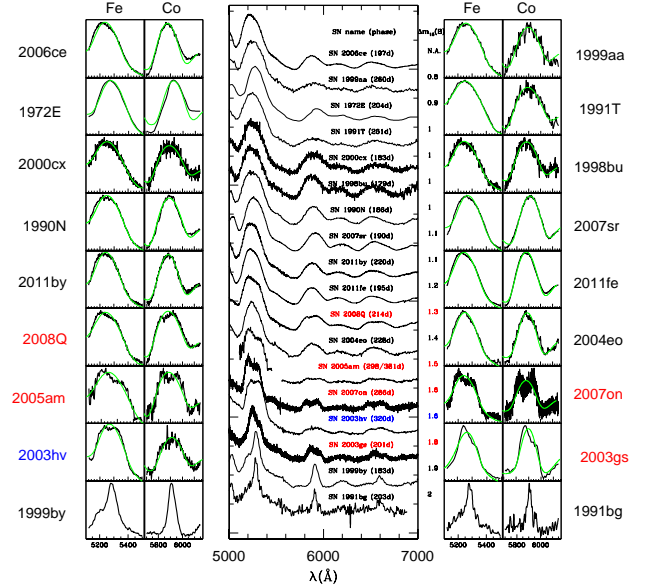}
\caption{20 SNe Ia with high-quality nebular spectra. They span the full range of $\Delta m_{15}(B)$ on the Phillips relation ($0.8<\Delta m_{15}(B)<2$) (middle panel, black). The spectra are sorted by the $\Delta m_{15}(B)$ values (shown on the right). Most of the spectra are single-peaked and are well fitted by the convolution of a simple quadratic velocity distribution with the template (SN 1999by). The best-fit single-peak models are shown in the sub-panels (green) where they are scaled to allow comparison with the corresponding spectral features (black) due to Fe and Co emissions, respectively. SN 1991bg is nearly identical to SN 1999by and is not fitted. Binned spectrum of SN 2003gs is shown in the right sub-panels to enhance clarity. Three SNe are identified with doubly-peaked profiles (SN 2007on, SN 2005am and SN 2003gs). SN 2008Q shows flat-top line profiles. SN 2003hv shows hints of departure 
from single-peak but is ambiguous to tell whether it has double-peak/flat-top profile. SN 2005am spectrum is taken from two epochs (298d and 381d) since the high SNR spectrum at 298d did not cover the $\sim 5300{\AA}$ feature.}
\end{figure}

\begin{figure}
\includegraphics[width=1.1\columnwidth]{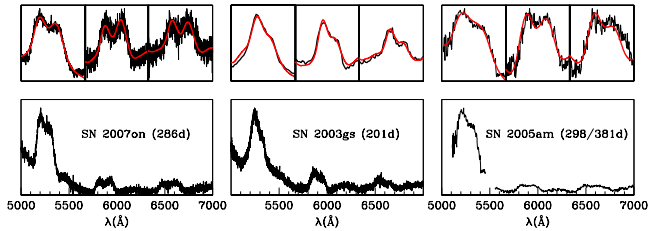}
\caption{The three SNe spectra that show double-peak profiles.  For each SN, a simple common velocity kernel with two quadratic components is used to fit all three features via convolutions with the template spectrum of SN1999by (see Fig. 1). The convolved spectra are in excellent agreement with all the features of SN 2007on, SN 2003gs and SN 2005am, suggesting that the underlying velocity distribution are bimodal for these SNe.}
\end{figure}

\subsection{Bi-modal Velocity Distributions Are Common, Especially  Among Low-Luminosity SNe}

{The sample of 20 SNe with high-SNR nebular spectra are shown in Figure 3 in ascending order of $\Delta{m_{15}(B)}$ (if available). To identify doubly-peaked lines, we fit each spectrum using a simple, single-peak velocity convolution kernel with $dM/dv_{\rm los}\propto {\rm max}[1-(v_{\rm los}/v_{\rm mod})^2,0]$.  The Fe and Co features of many SNe can be well fitted by single-peaks as shown in Figure 3. The use of SN 1999by(/1991bg) as template(s) is justified mainly on an empirical
ground, and it could be problematic if they result in different explosion physics from 
other SNe Ia, as argued in some works (e.g., \citealt{mazzali12}). 
Nevertheless, as discussed in Sec 2.2, we primarily use the  clean [CoIII] 5900\AA\, feature
for double-peak identification, which does not depend on the convolution method.  
 We identify 3 SNe, SN 2007on, SN 2005am and SN 2003gs, with clearly doubly-peaked spectra that consistently appear in the three features (marked in red).  Figure 4 shows that all three spectra can be well fitted by convolving a two-component velocity profile with the template.

SN 2008Q shows a ``flat-top'' profile that appears in all three Fe/Co features (with low-level
of NaI D absorption at $\sim 5900\AA$). SN 2003hv $(\Delta{m_{15}(B)}=1.6)$  shows hints of a doubly-peaked or a flat-top profile, but the data do not allow unambiguous identification. A clear flat-top profile was reported for SN 2003hv for [Fe II] at 1.644 micron, and it was regarded as evidence
for a ``hole'' in the $^{56}$Ni distribution (see, e.g., \citealt{2003hv, holetheory}). SN 2003hv was also reported to have nebular line blue-shifted by $\sim 3000 {\,\rm km/s}$ \citep{maeda}. The flat-top profile and the line shift could be the result of 
an underlying bi-model velocity distribution, but unlike a double-peak profile, 
flat-top profile does not allow unambiguous identification of bi-modality.

Unquantified selection effects exist in this sample, so the numbers above are not suitable for statistics. Nevertheless, the SNe Ia with underlying bi-modal velocity distributions are definitely not rare.  It is interesting to note that, all the SNe with evidences for bi-modal velocity distributions have relatively large $\Delta{m_{15}(B)}$: $\sim 1.8$ (SN 2003gs), $\sim 1.6$ (SN 2007on), 
$\sim 1.5$ (SN 2005am) and $\sim 1.3$ (SN 2008Q). The bi-modal velocity distribution must be quite common among these fast declining, low-luminosity SNe with $\Delta{m_{15}(B)}>1.3$, 
which comprise $\sim 40\%$ of all SNe Ia in a volume-limited sample \citep{licomplete}.} 

\section{Theoretical Implications and Discussion}
A bi-modal velocity distribution is naturally expected from direct WD-WD collisions due to the detonations of both WDs, which occur for all cases in the high-resolution 2D simulations for zero impact-parameter WD-WD collisions \citep{kushnir13}. However, bi-modal $^{56}$Ni velocity distributions are rare in these 2D simulations. Figure 5 shows the results of a 3D simulation of the collision of two $0.64 M_{\odot}$ WDs with a non-zero impact parameter of 0.2. The impact parameter is defined as the ratio between $r_p$, the minimal separation between the WDs along their trajectory that would have been obtained if they were point masses, and the sum of the two WD radii. The simulation is performed using FLASH 4.0 \citep{dubey, flash2} with a 13 isotope alpha-chain reaction network \citep{timmes} and a $8\rm km$ resolution, comparable to the converged 2D resolutions in \citet{kushnir13}. The upper panel of Figure 5 shows the projected velocity distribution of the total ejecta mass and the $^{56}$Ni mass in the WD-WD orbital plane. The $^{56}$Ni mass consists of two components separated by several thousand km/s. The bottom panel shows the LOS $^{56}$Ni distributions for numerous viewing angles. The chance of seeing doubly-peaked line profiles with similar velocity separations as the observed ones is significant. It is worth noting that a narrow velocity distribution similar to SN 1999by/1991bg is observed from directions perpendicular to the line connecting the centers of the two $^{56}$Ni components. 

{It is important to note that, even though the $^{56}$Ni bi-modality are commonly 
expected from WD-WD collision models, not all WD-WD collisions result in bi-modality.  As suggested by the results of 2D simulations, collisions with zero impact-parameter rarely produce
bi-modality. Besides the impact parameter, the masses
of the WDs, and in particular their mass ratio, can conceivably play an important role in determining 
bi-modality. 3D simulations with the full range of impact parameters and WD masses are required to derive statistics of the expected velocity distributions (Kushnir et al., in prep).}

\begin{figure}
\includegraphics[width=\columnwidth]{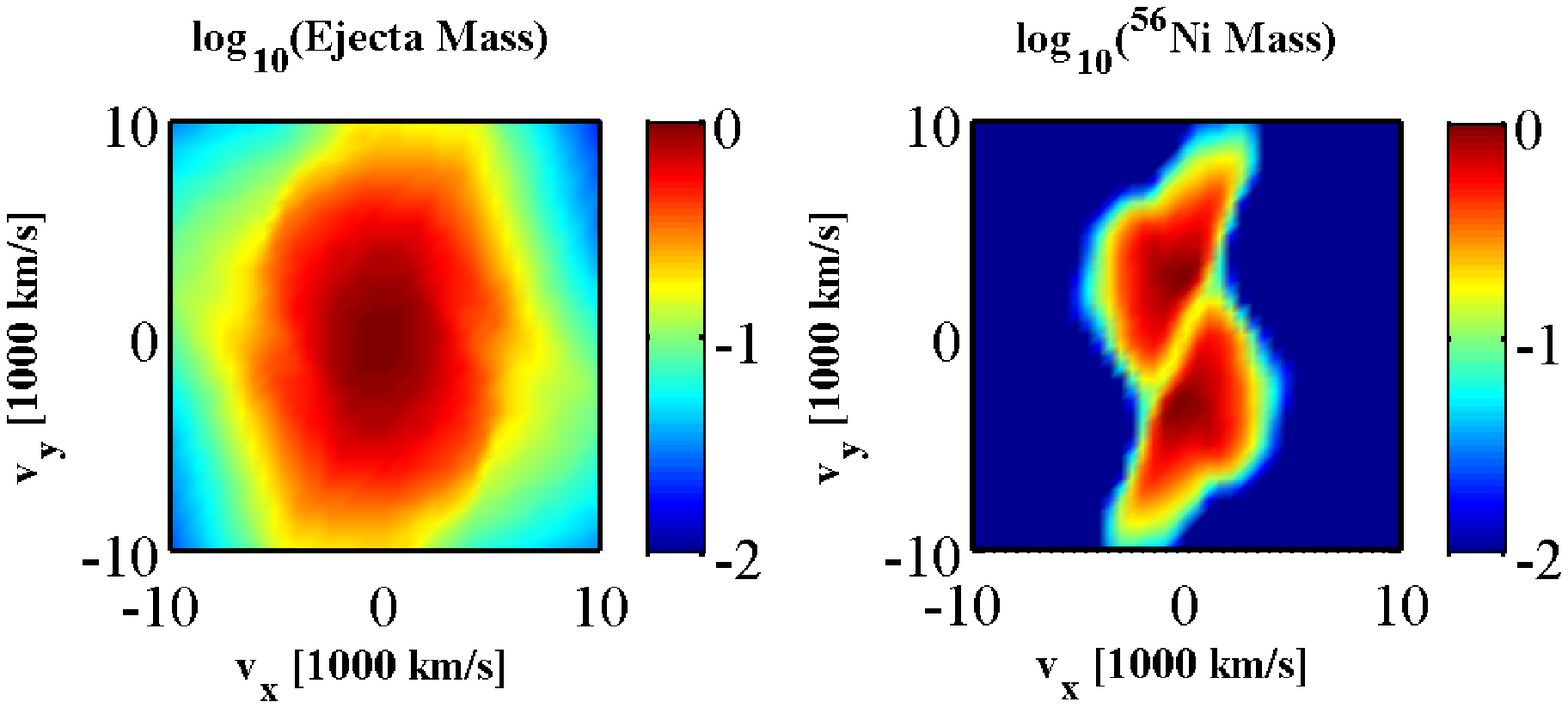}
\includegraphics[width=\columnwidth]{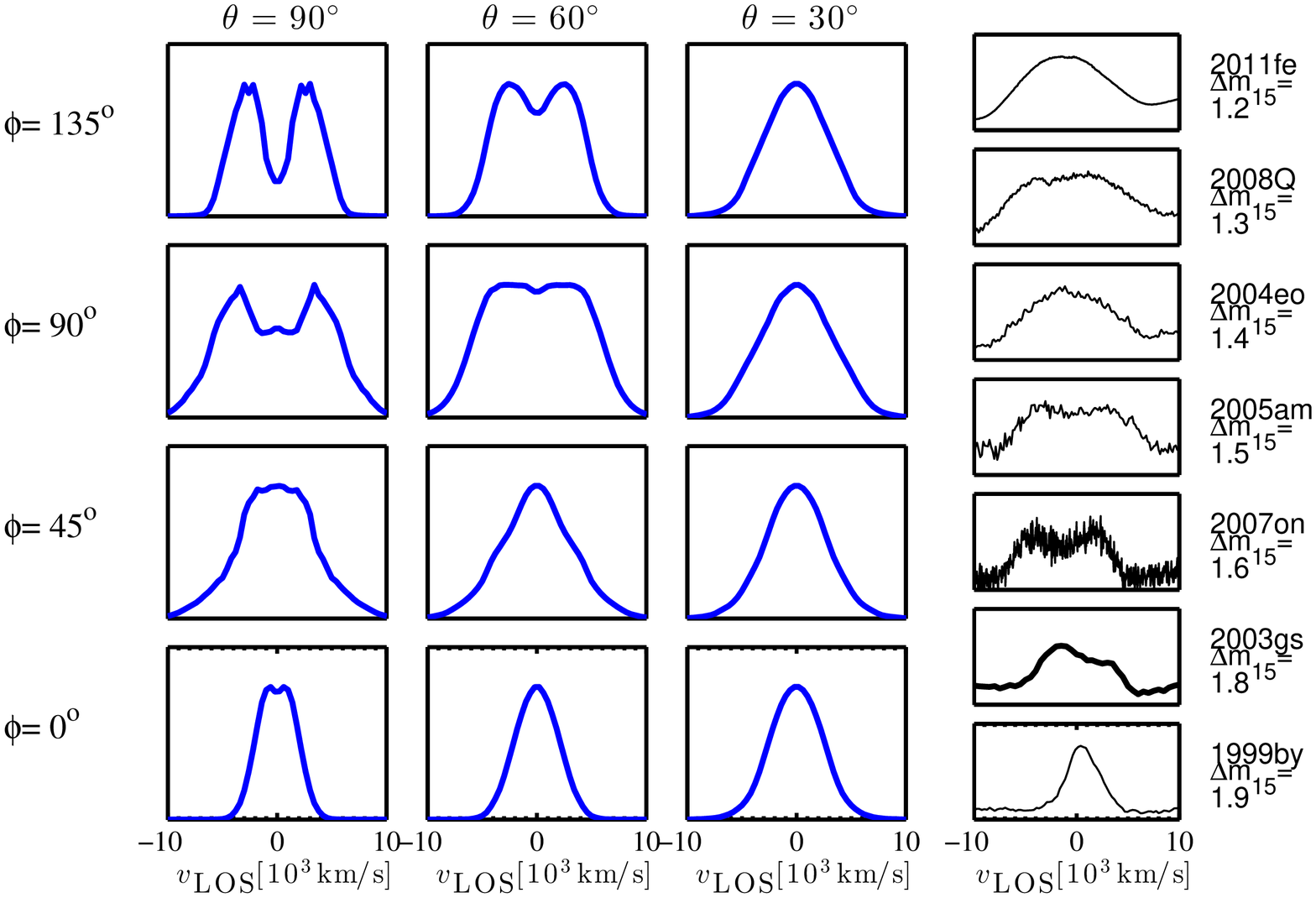}
\caption{Upper Panels: Ejecta at late, homologously-expanding phase from a 3D simulation of a $0.64M_{\odot}$-$0.64M_{\odot}$ WD collision with an impact parameter of 0.2. The projected velocity distribution of the $^{56}$Ni mass and total mass in the orbital plane are shown in the top right and left
panels. $^{56}$Ni mass is concentrated in two well separated components, which are related to the two ignition spots in the two WDs. Bottom Left: the resulting line of sight velocity distributions that are observed at numerous viewing angles (blue). The viewing direction is described by spherical coordinates with $\theta$ the polar angle with respect to
the z axis (direction of the angular momentum) and $\phi$ is the azimuthal angle in the $x-y$ plane with respect to the $x$ axis. Bottom Right: the resulting velocity distribution can be directly compared to the line profiles near $5900$\AA\,\,, which is a ``clean'' [CoIII] line with minor blend at the red side. The observations are shown in velocity space, $v=c(\lambda-5900{\AA})/5900$\AA\,\, for several SNe shown in Fig. 3.}
\end{figure}

Our results imply that any proposed scenario to explain the majority of SNe Ia must have a considerable fraction of the explosions producing bi-modal velocity distributions of $^{56}$Ni. Direct collisions of WDs is a promising channel to explain the bi-modality, and the $^{56}$Ni distribution can be definitively calculated with no free parameters from the collision model. High-quality nebular-phase spectra, emphasizing SNe with fast post-peak decline $\Delta{m_{15}(B)}>1.3$, are needed to quantify the occurrence rate of the doubly-peaked profiles. Bi-model velocity distributions can sometimes masquerade as single-peak line profiles, and for some of them, their bi-modal nature may be revealed by detecting the shifts of line peaks with respect to the rest frame.  {The continuum polarization 
measurements of SNe Ia at early epochs were found to be $\lesssim 1\%$ and the inferred
maximum departure from spherical symmetry was suggested to be $\lesssim 15\%$ (see e.g., \citealt{polarizationaraa, 2005ke}). 
Whether the bi-modality of $^{56}$Ni suggested in the nebular-phase spectra and expected from the collision models are consistent with the polarization measurements is an open issue, and resolving it will require radiation transfer calculations.} By comparison of accurate computations and comprehensive observations in the near future, the collision models can be unambiguously tested as the primary channel for type Ia SNe. The rich and detailed structure of the nebular line profiles will either be a smoking gun of the collision model or rule it out.

\section*{Acknowledgements}

{
We thank Avishay Gal-Yam, Andy Gould and the reviewer Mark Phillips for helpful comments.
We are grateful to Ben Shappee for providing the spectral files for SN 2011fe
 and Jeffrey Silverman for help with BSNIP. We thank the Carnegie Supernova Project, 
 Berkeley Supernova Ia Program, CfA Supernova Group for making their data public and 
 the Online Supernova Spectrum Database (SUSPECT) for collecting a comprehensive set of 
 archival SNe spectra -- without these efforts, this work would have not been possible. 
 S.D. is supported by ``the Strategic Priority Research Program-The Emergence of Cosmological 
 Structures'' of the Chinese Academy of Sciences (Grant No. XDB09000000). D. K. gratefully 
 acknowledges support from Martin A. and Helen Chooljian Founders' Circle. 
 Support for J.L.P. is in part provided by FONDECYT through the grant 1151445 and by the Ministry
 of Economy, Development, and Tourism's Millennium Science Initiative through grant IC120009, 
 awarded to The Millennium Institute of Astrophysics, MAS.
 FLASH was in part developed by the DOE NNSA-ASC OASCR Flash Center at the University of 
 Chicago. Computations were partly performed at PICSciE and IAS clusters. 
This work used the Extreme Science and Engineering Discovery Environment (XSEDE), which is supported by NFS grant ACI-1053575.}
%%%%%%%%%%%%%%%%%%%%%%%%%%%%%%%%%%%%%%%%%%%%%%%%%%



%\bibliographystyle{mnras}


%%%%%%%%%%%%%%%%%%%%%%%%%%%%%%%%%%%%%%%%%%%%%%%%%%

%%%%%%%%%%%%%%%%% APPENDICES %%%%%%%%%%%%%%%%%%%%%

\newpage
\appendix
\section{The Convolution Kernel for the Bi-modal Velocity Distribution}

In Figure 4, the velocity convolution kernel 
to fit doubly-peaked profiles is given by
\begin{equation}
\frac{dM}{dv_{\rm LOS}}\propto P_1 + r\times P_2, {\rm where}\, P_1 = {\rm max}\left(1-\frac{(v_{\rm LOS}-v_{\rm shift,1})^2}{v_{\rm mod,1}^2}, 0\right) \nonumber
\end{equation}
\begin{equation}
{\rm and}\, P_2 = {\rm max}\left(1-\frac{(v_{\rm LOS}-v_{\rm shift,2})^2}{v_{\rm mod,2}^2}, 0\right).
\end{equation}
There are 5 free parameters: two shifts $v_{\rm shift,1,2}$, the two widths $v_{\rm mod,1,2}$ and the peak ratio of the components, $r$. The shifts can be alternatively described as a velocity shift $v_{\rm shift}=0.5(v_{\rm shift,1}+v_{\rm shift,2})$ and velocity separation $v_{\rm sep}= v_{\rm shift,2}-v_{\rm shift,1}$. The best-fit parameters $(v_{\rm shift}, v_{\rm mod,2}, v_{\rm mod,2}, v_{\rm sep}, r)$
are $(-2109, 1338, 2000, 5330, 0.89)$, $(1132, 522, 711, 4100, 0.54)$ and 
$(1111, 3347, 4195, 6500, 0.91)$ for SN 2007on, 2003gs and 2005am, respectively.

\section{On the choice of the template and spectral features for convolution}

\begin{figure}
\includegraphics[width=\columnwidth]{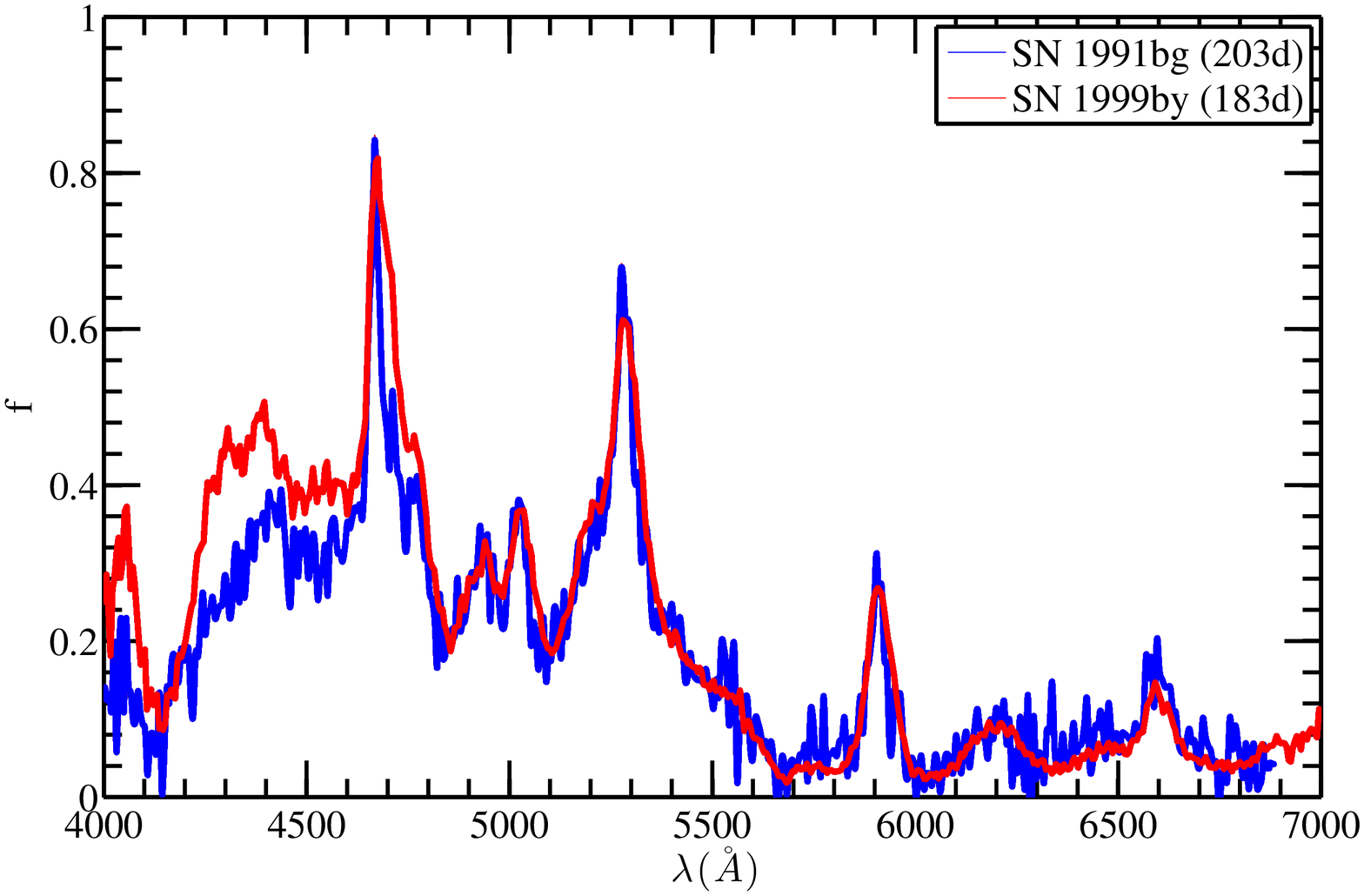}
\caption{Nebular-phase spectra for SN 1991bg (blue) and SN 1999by (red) normalized 
in flux for comparison.}
\end{figure}

\begin{figure}
\includegraphics[width=\columnwidth]{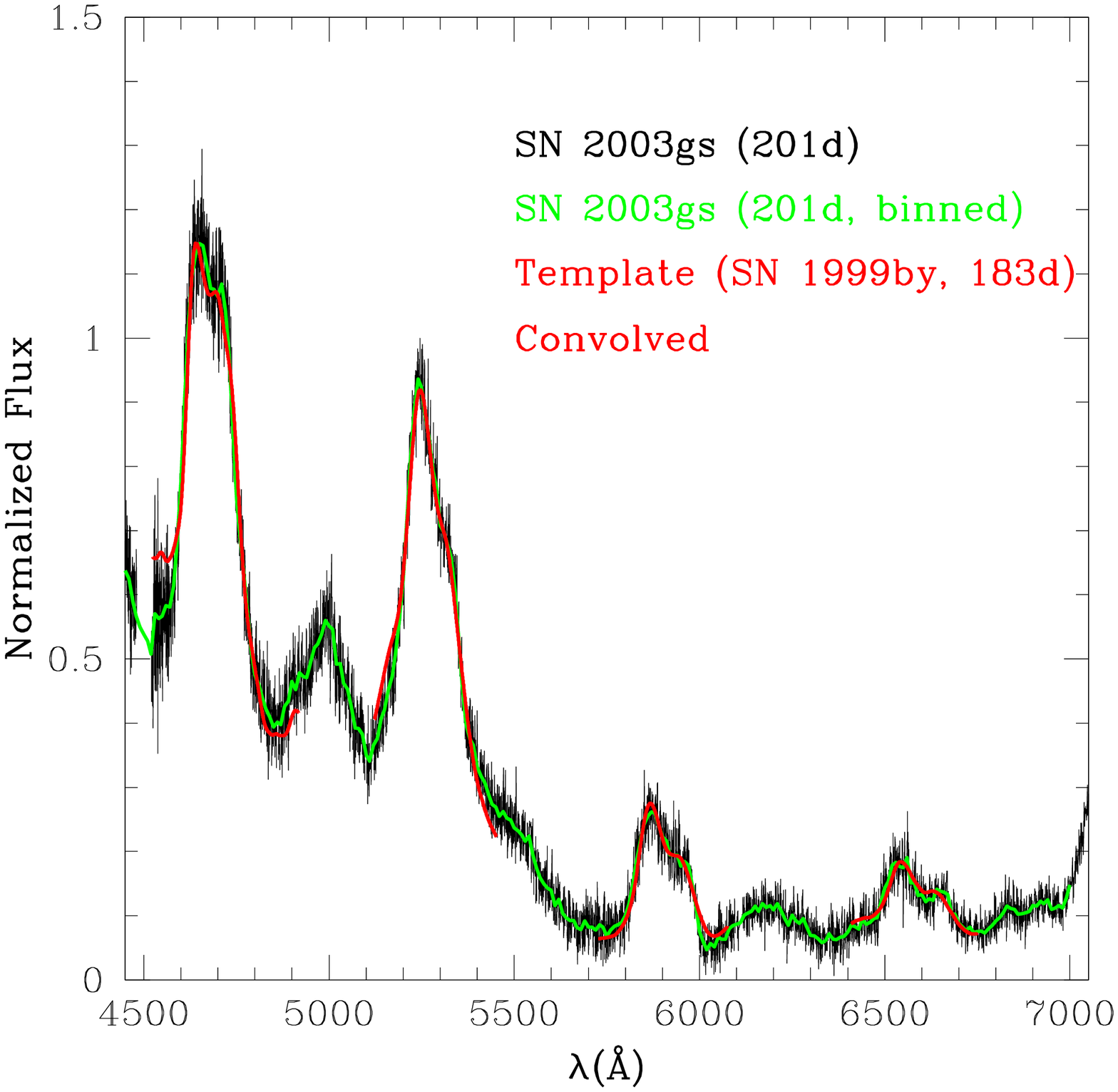}
\includegraphics[width=\columnwidth]{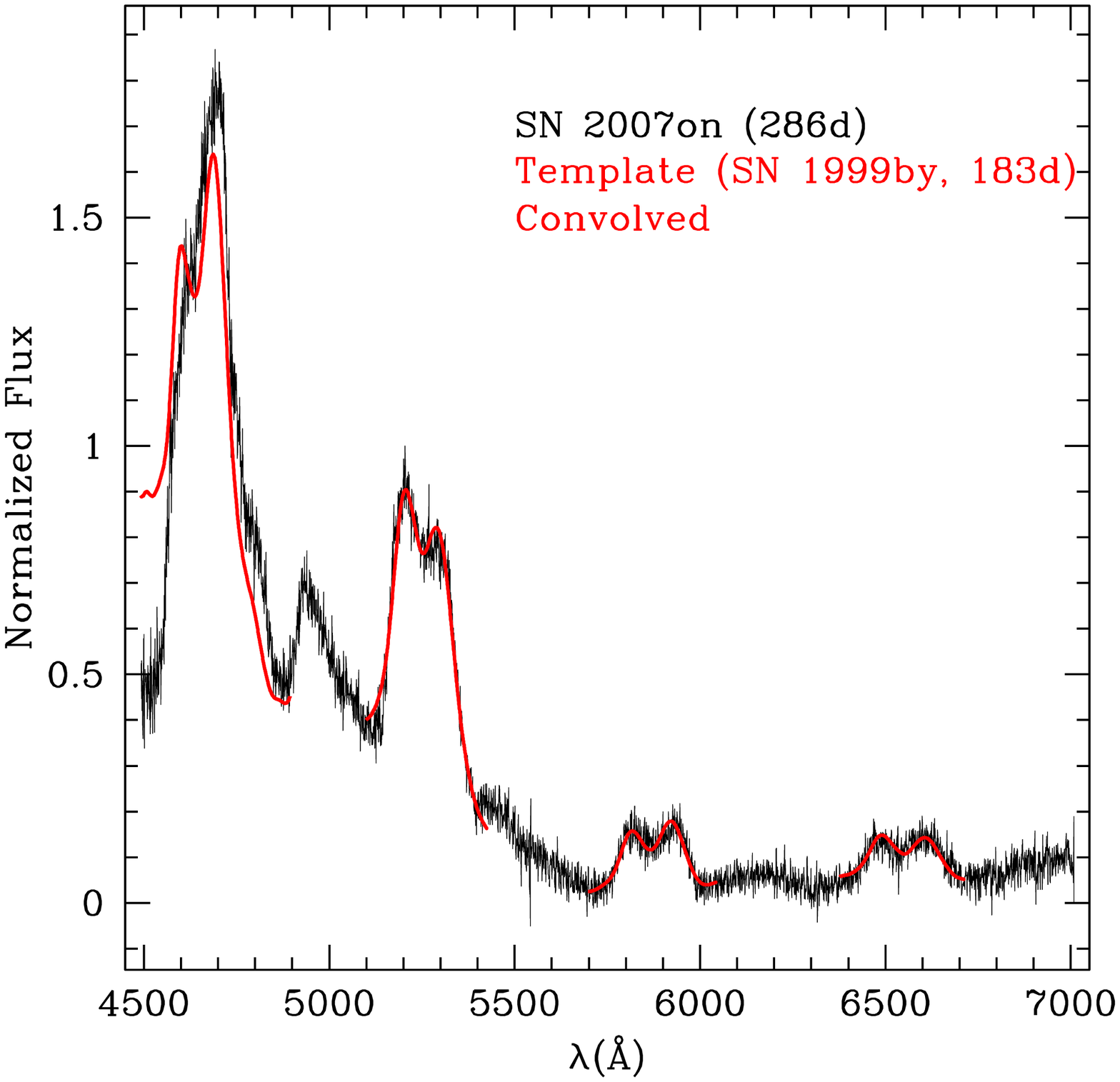}
\caption{The convolutions of the bi-modal velocity kernels (Appendix A)
of the template SN 1999by (183\,d) are extended to the $\sim 4700\,\AA$ feature, 
and the convolved spectra are shown in red. 
The results for SN 2003gs (201\,d) are shown in the upper panel 
(black for the un-binned and green for the binned spectrum to enhance display) 
and SN 2007on (286\,d) shown in the lower panel (black). 
The convolved spectra are in good agreement with SN 2003gs, which was 
observed at a similar phase to the template spectrum of 1999by, while the 
agreement is not good for SN 2007on, which was observed at a much later 
phase than SN 1999by. For fitting each spectrum, the velocity kernel is common 
to all spectral features while the normalization and baseline 
are allowed to vary freely for each feature. For SN 2003gs, we also test the 
convolution of 
the entire SN 1999by template spectrum with the bi-modal velocity kernel (rather than fitting individual spectral regions), 
and the convolved spectrum shows remarkable agreement with 
that of SN 1999by across the entire spectrum.}
\end{figure}

The direct convolution method applied in this work makes use of multiple 
nebular-phase Co and Fe spectral features to study the underlying 
velocity profile of $^{56}$Ni in the ejecta. Among all nebular-phase 
spectra in our collection, the spectra of SN 1999by and SN 1991bg 
have the narrowest line widths, which make them suitable as the 
templates for convolution. The spectra of SN 1999by (183 d) and 
SN 1991bg (203 d) are plotted in Fig. B1. For the work presented 
in the main text, we used the spectrum of SN 1999by as the template 
due to relatively high SNR compared to that of SN 1991bg.

The ideal nebular spectral features to use shall have little blending 
from nearby spectral lines and exhibit good consistency in line 
shapes for different SNe. The shapes of spectral features in the 
range of $\sim 5000$ \AA\ and $\sim 7000$ \AA\ are essentially 
identical between the two SNe. The [CoIII] feature at $5900$ \AA\ is 
especially clean with its wings falling close to zero at both blue and 
red ends. Moreover, theoretical study of nebular-phase spectrum 
by \citet{thesis} shows that the $5900$ \AA\ [CoIII] 
feature is devoid of blending from any other lines. The Fe feature 
at $\sim5300$ {\AA} (due to the blend of [FeIII] and [FeII]) and the 
[CoIII] feature at $\sim6600$ {\AA} (blended with [FeII]) are less clean, 
and they are used in the study only to aid the line profile analysis 
mainly based on the $5900$ {\AA} feature.

In contrast, the spectra features in the range of 
$\sim 4000 - 5000$ \AA\ show significant difference in shape between 
SN 1999by and SN 1991bg, in particular for the prominent Fe feature at 
$\sim 4700$  \AA\, which is frequently used in many studies of SNe Ia nebular spectra. 
The $\sim 4700$  \AA\ feature seems to be significantly affected 
by complicated blends of neighbouring lines. 
This makes this feature unsuitable for direct template convolution.

Nevertheless, it is instructive to examine the $\sim 4700$ \AA\  
features for the supernovae with bi-model velocity distribution 
deduced from the $\sim 5900$ \AA\ features. In our sample, 
the nebular-phase spectra for SN 2005am do not possess the 
region near $\sim 4700$ \AA. The spectrum of SN 2003gs is 
the more promising case to be modelled by convolution as it is 
taken at 201\,d, similar to the phase of the template SN 1999by spectrum (183\,d).
The upper panel of Fig. B2 shows the convolution of 
the SN 1999by template spectrum with the bi-modal velocity 
convolution kernel given in Appendix A (i.e., extending 
the convolutions shown in Figure. 4 
in the main text to the $\sim 4700$ \AA\ feature). There is a remarkable agreement in 
the line shape at $\sim 4700$ \AA\  
between the convolved spectra and that of SN 2003gs.
The nebular-phase spectrum of SN 2007on is taken at 286\,d, much 
later than the template spectrum at $\sim 200$\,d. 
In the lower panel of Fig. B2, we compare the 
spectra between SN 2007on taken at 286\,d and the convolved 
SN 1999by spectra (183\,d), and in this case 
the agreement is not good for the $4700$ \AA\ feature. 
This is probably due to differences in the complex blend in this region between the two supernovae (SN 1999by and 2007on) and/or at the different phases.
% In addition, the difference in physical 
%states of ejecta between the two supernovae may also play a role. 
%We also compare the $4700$ \AA\ features between SN 2007on (286 d) 
%and SN 2011fe taken at a very similar epoch (275 d), and they 
%show striking differences, which are expected for different underlying velocity 
%distributions as indicated by the $5900$ \AA feature.

Despite the complexity of the $4700$ \AA\ feature demonstrated above, 
the case of SN 2003gs suggests that there is some regularity within 
this spectral region that can be explored. 
For SN 2003gs, we also convolve
the whole SN 1999by template spectrum with the bi-modal velocity kernel (rather than fitting individual spectral regions), 
and the convolved spectrum agrees remarkably well with 
that of SN 1999by across the entire spectrum.
Observationally, a larger sample of high 
quality nebular spectra spanning a larger 
range of epochs beyond 200\,d, in particular for the 1991bg-like events, would
help greatly to  elucidate the situation empirically. Further theoretical works 
would be needed explain the difference in line profiles for the $4700$ \AA\ 
feature between SNe (such as SN 1991bg and 1999by). The 
velocity profile deduced from the $5900$ \AA\ can  
be helpful in the modeling endeavours to interpret the more complex 
nebular lines such as the $4700$ \AA\ feature.


%%%%%%%%%%%%%%%%%%%%%%%%%%%%%%%%%%%%%%%%%%%%%%%%%%

% Don't change these lines
\bsp	% typesetting comment
\label{lastpage}
\end{document}